\documentclass[12pt]{iopart}
\usepackage{graphicx} 
\usepackage{subfigure}

\begin{document}

\title{Wirebond crosstalk and cavity modes in large chip mounts for superconducting qubits}

\author{J Wenner, M Neeley\footnote{Present address: Lincoln Laboratory, Massachusetts Institute of Technology, Lexington, MA 02420, USA}, Radoslaw C Bialczak, M Lenander, Erik Lucero, A D O'Connell, D Sank, H Wang\footnote{Present address: Department of physics, Zhejiang University, Zhejiang 310027, China}, M Weides\footnote{Present address: National Institute of Standards and Technology, Boulder, Colorado 80305, USA}, A N Cleland and John M Martinis}
\address{Department of Physics, University of California, Santa Barbara, CA 93106, USA}
\ead{martinis@physics.ucsb.edu}

\begin{abstract}
We analyze the performance of a microwave chip mount that uses wirebonds to connect the chip and mount grounds. A simple impedance ladder model predicts that transmission crosstalk between two feedlines falls off exponentially with distance at low frequencies, but rises to near unity above a resonance frequency set by the chip to ground capacitance. Using SPICE simulations and experimental measurements of a scale model, the basic predictions of the ladder model were verified. In particular, by decreasing the capacitance between the chip and box grounds, the resonance frequency increased and transmission decreased. This model then influenced the design of a new mount that improved the isolation to -65\,dB at 6\,GHz, even though the chip dimensions were increased to 1\,cm by 1\,cm, 3 times as large as our previous devices. We measured a coplanar resonator in this mount as preparation for larger qubit chips, and were able to identify cavity, slotline, and resonator modes.
\end{abstract}

\pacs{06.60.Ei, 85.25.Am, 84.40.Dc, 84.30.Bv, 03.67.Lx}
\submitto{\SUST}

\section{Introduction}

One promising approach toward quantum computation employs superconducting qubits \cite{Clarke2008}. Such qubits have been used to demonstrate simple gates \cite{Yamamoto2003,Plantenberg2007,DiCarlo2009,Yamamoto2010} and quantum algorithms up to three qubits \cite{Neeley2010,DiCarlo2010}. As additional qubits are added, they will require increasing room on each chip, eventually necessitating a larger chip and increasing numbers of microwave control lines \cite{Mariantoni2010}. It will thus become crucial to carefully engineer the mounting box, both because the increased density of input lines could lead to greater microwave crosstalk between the lines, and because stray cavity modes, which can present new modes for dissipation \cite{Schuster2010}, can appear with larger mounts.

For superconducting microwave devices such as qubits and kinetic inductance detectors \cite{Mazin2009}, most chip mounts have placed or glued a chip onto a metal base \cite{Cicak2009,Deppe2007,Mazin2010,Frunzio2005,PalacoisLaloy2008}. Wirebonds are then used to connect the chip ground plane and pads to the external mount's ground plane and feedlines, which are typically made from a circuit board.

In this paper, we show that although such a mount may work well for the current generation of small chips, increasing the chip size will cause the ground connection to fail at microwave frequencies. Specifically, crosstalk will increase because of increased capacitance from the chip to mount ground, contrary to the usual assumption that tying the grounds together with capacitance improves their connection. With a simple ladder model, we explicitly show this capacitance introduces a traveling wave mode between the chip and mount grounds, thereby giving a continuum of resonance modes which decouples the two grounds and provides a mechanism for large crosstalk.

To understand the effects from the wirebonds and mount, we first develop a semiquantitative model, and then discuss experimental tests using a simple scaled-up system. The predictions of this model were then used to design a new sample mount with a reduced capacitance between the mount and chip ground planes. An actual resonator device was then measured for this design. This mount accommodates a chip three times the area of our current generation of qubits; with this increase in size, and with added cavities to reduce the capacitances between the box and chip ground planes, we identified sources of cavity and chip modes up to 20\,GHz.

\section{Wirebond Crosstalk}

\subsection{Circuit Model}

\begin{figure}
\begin{center}
\includegraphics[width=3in]{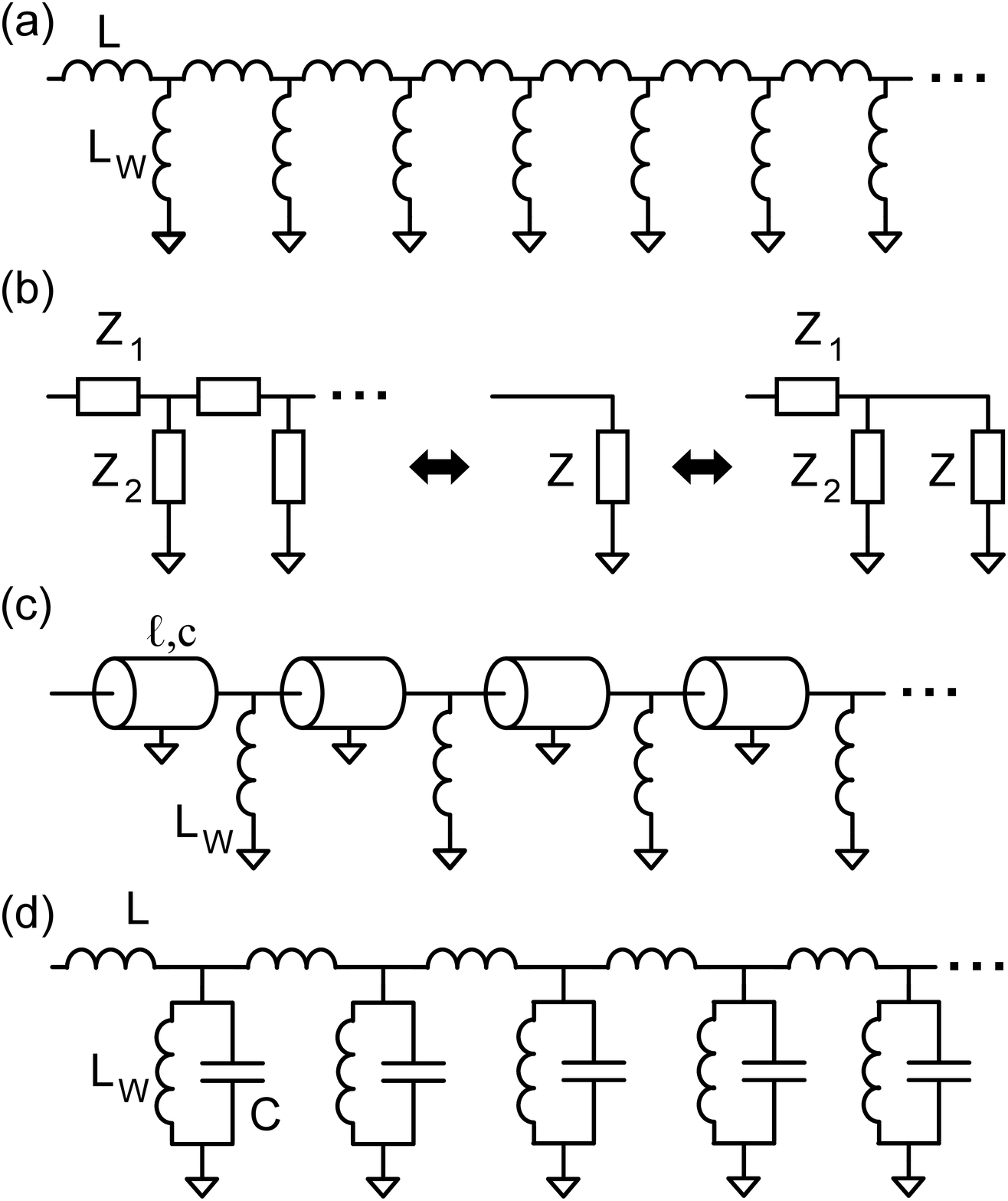}
\end{center}
\caption{Ladder model of chip and wirebonds. (\textbf{a}) The chip ground plane and wirebonds are modeled as an inductor ladder. (\textbf{b}) A ladder with impedances $Z_1$ and $Z_2$ can be replaced by an effective impedance $Z$, or equivalently, by a single rung in parallel ($||$) with the rest of the ladder, which is again equal to $Z$. Equating the latter two cases gives a quadratic equation for the effective impedance: $Z = Z_1 + (Z_2 || Z)$. This can be solved for $Z$, and then the single rung is a voltage divider, giving $V_1/V_0 = (Z_2 || Z) / (Z_1 + (Z_2 || Z))$ \cite{Feynman}. (\textbf{c}) Schematic of the ladder model that incorporates the distributed nature of the edge inductance and ground capacitance. (\textbf{d}) Treating the ground capacitance as lumped elements adds a parallel impedance of opposite sign to the wirebond inductance, leading to an open circuit at the resonance frequency $f_{res} = 1/2\pi\sqrt{L_W C}$.}
\label{fig:schematic}
\end{figure}

When mounting a microwave device, it is necessary to ensure that the chip is well grounded to the mounting box. This may be achieved using many short, closely-spaced wirebonds around the edge of the die. To understand the effect of this network, we consider the simple model illustrated in Figure \ref{fig:schematic}, which will later be experimentally verified. Here, the grounding wirebonds between an input and an output feedline are considered to be an impedance ladder, with a node at each point where a wirebond connects to the chip ground plane.

Each wirebond can be modeled as a wire having inductance $L_W$ between each node and the box ground.  This inductance is proportional to the length of the wirebond, with a proportionality constant of approximately 1\,nH/mm for typical wirebonds due to ground plane effects \cite{Grover1912}. The length of the wirebond can be modeled to be approximately the minimum possible length for the wirebonds, the distance $g$ between the chip and the edge of the mount. In addition, the edge of the chip ground plane gives an inductance $L$ between two adjacent nodes.  Its magnitude is $L = \ell d$ for an edge inductance per unit length $\ell$ and distance $d$ between adjacent wirebonds.  We assume the corresponding edge inductance from the mount ground plane is negligible because adjacent nodes are connected through the bulk. These two types of inductances give rise to a ``ladder model'' for the interface between the chip and mount ground planes, as illustrated in Figure \ref{fig:schematic}(a).

We model microwave crosstalk as coming from voltage propagating along the ground nodes on the chip.  The crosstalk between two ports can then be solved geometrically by following the approach \cite{Feynman} illustrated in Figure \ref{fig:schematic}(b), where the impedance looking into the ladder is given by $Z$ and is independent of node position.  The ratio of voltages between adjacent nodes of the inductor ladder is
\begin{eqnarray}
\frac{V_1}{V_0} = \left[1 + \frac{\zeta}{2}\left(1 +
\sqrt{1+\frac{4}{\zeta}}\right)\right]^{-1}
\label{eq:ladder}\\
\zeta = \frac{Z_1}{Z_2}=\frac{L}{L_W}.
\end{eqnarray}

\begin{figure}
\begin{center}
\includegraphics[width=3in]{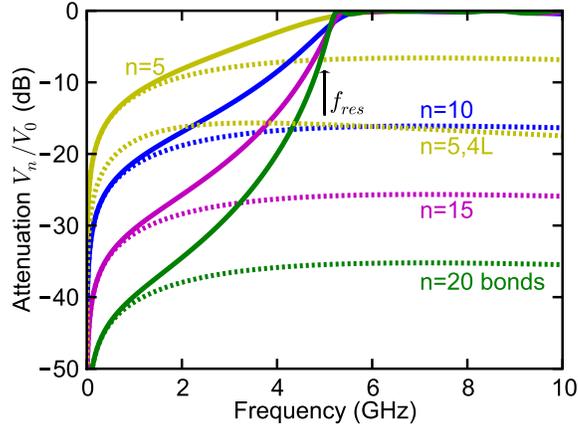}
\end{center}
\caption{SPICE simulation of the ladder model.  We plot voltage attenuation $V_n/V_0$ versus frequency for varying numbers of wirebonds $n = 5$ (yellow), 10 (blue), 15 (purple) and 20 (green).  Dashed lines are for the inductor ladder, whereas the solid lines include capacitance. The parameters are similar to those in model A of Table\,\ref{table:measurements} for the 10x scale model, with $L = 0.5$\,nH, $L_W = 2.54$\,nH and $C = 0.4$\,pF, giving $L/L_W = 0.2$. We observe greater attenuation for increasing bond numbers both with and without capacitance. However, the transmission approaches unity above the resonant frequency of $f_{res} = 1/2\pi\sqrt{L_W C} = 5.0$\,GHz when capacitance is included. The lower yellow dashed trace is for 5 bonds without capacitance, but with $L=2$\,nH to emulate the same distance as for the case of 20 bonds but with 1/4 the bond number; this result demonstrates the advantage of adding additional bonds rather than increasing the inductance ratio $L/L_W$ with greater spacing.}
\label{fig:spice}
\end{figure}

As the inductance ratio $L/L_W$ increases, the voltages along the nodes are more quickly suppressed. This implies that $L_W$ should be reduced as much as possible by keeping wirebonds short. The voltage ratio between widely spaced points is given by a geometric sequence $V_n/V_0 = (V_1/V_0)^n$. Note that while increasing the distance $d$ between bonds increases $L/L_W$ and thus reduces $V_1/V_0$, it also reduces the number of bonds $n$. As illustrated in Figure \ref{fig:spice}, for increased attenuation it is more advantageous to have a high density of wirebonds than to space fewer wirebonds over the same length of ground plane.

Capacitance between the ground plane of the chip and the mount must also be included in the model.  For a chip placed on a metal plane, this capacitance is distributed to all points on the ground plane of the chip, but for simplicity we treat it as being connected along the edge of the chip. This capacitance, along with the inductance $L$, is distributed over the length between wirebonds, giving a transmission line as illustrated in Figure \ref{fig:schematic}(c). By employing a $\pi$ model \cite{Johnson}, the capacitance can be divided between the wirebond nodes as lumped elements. As with parallel plate or coplanar capacitors, this capacitance $C$ is proportional to the distance $d$ between adjacent wirebonds, giving $C=cd$ for capacitance per unit length $c$. With these assumptions, the chip mount is represented by the model illustrated in Figure \ref{fig:schematic}(d), where capacitance is placed in parallel to the bond-wire inductance. The ratio of voltages between adjacent nodes can be solved in the same manner as for (\ref{eq:ladder}), but here
\begin{equation}
\zeta = \frac{L}{L_w}-\omega^2LC.
\end{equation}
When adding capacitance, the important result is that the wirebond inductance resonates with the capacitance at a frequency $f_{res} = 1/2\pi\sqrt{L_W C}$, resulting in a loss of shunting from the wirebonds.  At (and above) this frequency, signals propagate along the chip ground without being strongly attenuated.

To account for the finite size of the ladder, we also simulated this circuit with SPICE for various numbers of bonds and ratios of inductance and capacitance. These simulations confirmed quantitatively the predictions of the simple model, particularly that the transmission approaches unity above $f_{res} = 1/2\pi\sqrt{L_W C}$ and increasing numbers of bonds attenuate voltage more strongly at frequencies below the resonance. Figure\,\ref{fig:spice} shows a typical simulation result, corresponding to model A of Table\,\ref {table:measurements}.

\begin{figure*}
\includegraphics[width=6in]{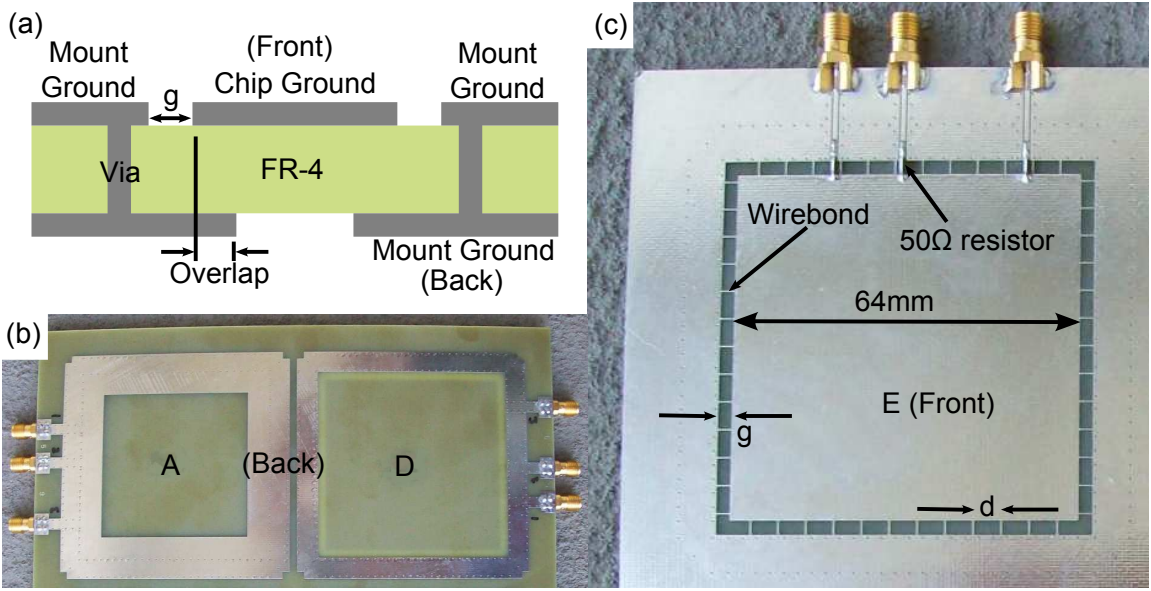}
\caption{Scale model of the chip mount. Each model is 64\,mm wide and high, which is 10 times larger than our standard chips. (\textbf{a}) Cross-section view (not to scale). The model consists of a chip ground (the center square on the front side) and two chip mount grounds (the two hollow squares on both the front and back sides), where the chip mount grounds are connected through vias. A gap in the front side models the separation between the chip and mount grounds. (\textbf{b}) View of the back of panels A and D demonstrate varying overlaps of the mount ground plane and the chip ground. (\textbf{c}) Front-side view of panel E. Each model has three microwave ports, connected to the chip ground via a 50\,$\Omega$ resistor, so that transmission can be measured for separation distances 13\,mm, 23\,mm, and 36\,mm. Wirebonds are modeled by traces 0.25\,mm wide connecting the chip and mount grounds.}
\label{fig:scalemodel}
\end{figure*}

We also checked the assumption of a lumped model for the ground inductance and capacitance by simulating short lengths of transmission lines, as illustrated in Figure \ref{fig:schematic}(c). By calculating the transmission matrix \cite{Pozar}, we solved for the voltage ratios between nodes and found that the resonance frequency $\omega_0/2\pi$ satisfies
\begin{equation}
1=\left(1+\frac{\sqrt{\ell / c} }{ 2 \omega_0 L_W } \right) \cos(\omega_0 d \sqrt{\ell c})
\label{eq:transline}
\end{equation}
When solved numerically, we found predictions changed less than 5\% as compared to the lumped model, indicating that the lumped model is adequate for the analysis presented here.

\subsection{Experimental Setup and Analysis}

\begin{table*}
\caption{\label{table:measurements}Parameters for scale model of chip mount. The first three are measured from the model geometry: $g$ is the gap between the ground plane and chip (the length of the bonds), $d$ is the spacing between bonds, and \textit{overlap} is the overlap distance between the chip ground plane (front side) and the mount ground plane (back). The capacitance $c$ and inductance $\ell$ per unit length are calculated using COMSOL, while the remaining parameters are calculated from $c$ and $\ell$. The lumped element (LE) resonance frequency is given by $f_{res}=1/2\pi\sqrt{L_W C}$ from the calculated inductance and capacitance. The transmission line (TL) resonance frequency is given by numerical evaluation of Eq.\,(\ref{eq:transline}).}
\begin{center}
\begin{footnotesize}
\lineup
\begin{tabular}{@{}llllllllllll}
\br
 & $g$ & $d$ & overlap & $c$ & $\ell$ & $L$ & $L_W$ & $L/L_W$ & $C$ & $f_{res}$ (LE) & $f_{res}$ (TL) \\
  & (mm) & (mm) & (mm) & (fF/mm) & (nH/mm) & (nH) & (nH) &  & (fF) & (GHz) & (GHz) \\
\mr
A & 2.5 & 2.5 & \m5.1 &  174.9 & 0.22 & 0.56 & 2.54 & 0.22 &  444 & \04.74 & \04.69 \\
B & 2.5 & 2.5 & \m2.5 &  114.6 & 0.29 & 0.74 & 2.54 & 0.29 &  291 & \05.85 & \05.78 \\
C & 2.5 & 2.5 & \m0.0 & \055.9 & 0.43 & 1.09 & 2.54 & 0.43 &  142 & \08.38 & \08.23 \\
D & 2.5 & 2.5 &  \0-2.5 & \032.4 & 0.58 & 1.47 & 2.54 & 0.58 & \082 &  11.0  &  10.75 \\
E & 2.5 & 5.1 & \m2.5 &  114.6 & 0.29 & 1.47 & 2.54 & 0.58 &  582 & \04.14 & \04.04 \\
F & 2.5 & 1.3 & \m2.5 &  114.6 & 0.29 & 0.74 & 2.54 & 0.15 &  145 & \08.28 & \08.23 \\
G & 5.1 & 2.5 & \m2.5 &  114.0 & 0.30 & 0.76 & 5.08 & 0.15 &  290 & \04.15 & \04.12 \\
H & 1.3 & 2.5 & \m2.5 &  117.2 & 0.28 & 0.71 & 1.27 & 0.56 &  298 & \08.18 & \08.00 \\
\br
\end{tabular}
\end{footnotesize}
\end{center}
\end{table*}

To verify semiquantitatively the predictions of this ladder model, we built a scale model of a chip, its wirebonds, and a mount, as illustrated in  Figure \ref{fig:scalemodel}. The scale models were commercially fabricated on FR-4 circuit board with double sided-copper and plated vias.  The effects of capacitance from the chip to the mount were modeled by varying amounts of metal overlap between the top and bottom sides of the board.  We assumed that the wirebond inductance scaled as 1\,nH/mm multiplied by the gap between the two ground planes, and that the edge inductance and capacitance scaled as the distance between adjacent wirebonds.

\begin{figure*}
\begin{center}
\includegraphics[width=6in]{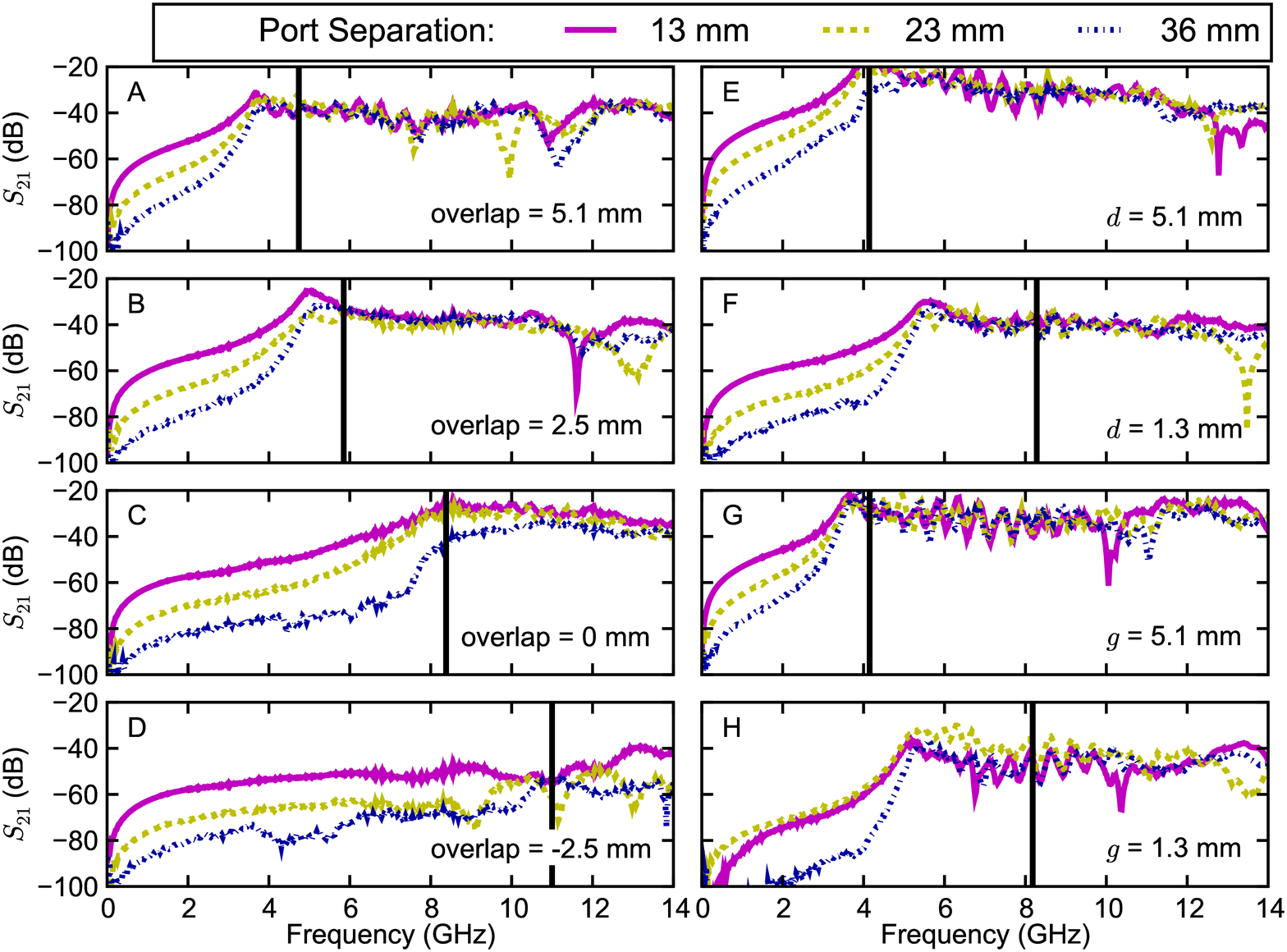}
\end{center}
\caption{Transmission versus frequency measured in 10x scale models.  Scaling implies frequencies would be 10 times larger for 6\,mm chips.  Data are plotted for three distances: 13\,mm (magenta solid), 23\,mm (yellow dashed), and 36\,mm (blue dash-dot).  At low frequency, the data show small transmission that increases with fewer wirebonds.  The transmission increases with frequency and peaks at a resonance frequency that is in qualitative agreement with the theoretical model. Panels A-D show a systematic increase in resonance frequency for decreasing chip capacitance to ground, which is in reasonable agreement with the prediction of $f_{res}$ indicated by the black vertical line.  The decrease in transmission at high frequencies is due to dissipation in the FR-4 and copper. Panels E and (F) show changes from the spacing $d$ between wirebonds, indicating longer (shorter) spacing produces a smaller (larger) resonance frequency, as expected.  Panels G and H show dependence on wirebond length $g$, which again change the resonance frequency and low-frequency transmission in the predicted direction. }
\label{fig:data}
\end{figure*}

On the top side, the chip was modeled by a 64\,mm metal square, ten times larger than our group's standard qubit chips. A gap of width $g$ separated this ground plane from that of the mount ground. Stretching across the gap were 0.25\,mm wide metal traces that simulated the wirebonds.  To measure the voltage attenuation down the chip, we included three signal leads, separated by 13\,mm and 23\,mmn, that were connected by a coplanar waveguide to a 50\,$\Omega$ resistor placed across the gap to the chip ground plane. The ground plane on the back side of the circuit board was connected with vias to the ground plane of the top mount.

\begin{figure}
\begin{center}
\includegraphics[width=3in]{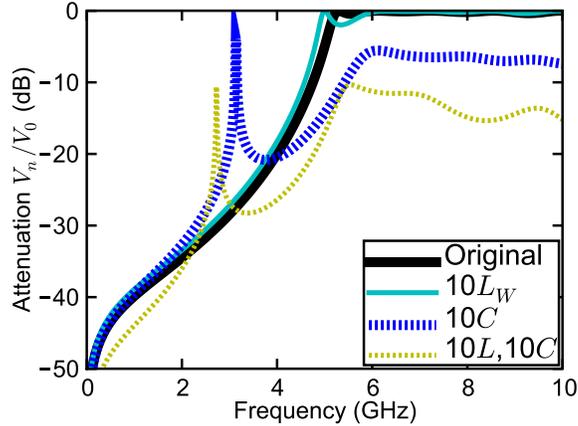}
\end{center}
\caption{SPICE simulation of nonuniformity in the ladder model. Parameters used are identical to Figure \ref{fig:spice} with 20 bonds, but with a change in $L$ and $C$ for one bond halfway down the line. The thick solid black trace is for an unchanged ladder, as in Figure \ref{fig:spice}. The thin dashed yellow trace corresponds to increasing both $L$ and $C$ by a factor of 10 to simulate a single gap lengthened by a factor of 10. For the thick dashed blue trace, $C$ is increased 10-fold to simulate unwanted capacitance with no gap in bonds. Finally, the solid cyan trace is for a 10-fold increase in $L_W$ to simulate a bond that is much too long.  While the discontinuities in capacitance lead to resonant transmission at low frequencies, an increased wirebond inductance does not, indicating that the grounding performance is less sensitive to irregularities in the wirebonds themselves.}
\label{fig:spice_res}
\end{figure}

As illustrated in Table\,\ref{table:measurements}, we used a default distance of 2.5\,mm for the gap between the front- and back-side ground planes, the distance between adjacent bonds, and the overlap distance between the front side chip and the back side ground plane. These three parameters were changed for the eight different models. For each model, the COMSOL multiphysics package \cite{Comsol} was used to calculate the edge inductance and capacitance per unit length; these parameters, along with the inductances and capacitances derived from these values, are also included in Table\,\ref{table:measurements}.

We used a vector network analyzer at room temperature to measure the scattering matrix for these scale models.  By measuring $S_{21}$ between pairs of the three connectors, we were able to determine attenuation for three different lengths.  The resulting traces are shown in Figure \ref{fig:data}.

As predicted by the frequency dependence of the model, we observed rolloff at very low frequency, then rising transmission until a resonance frequency, after which the transmission remains high.  Below the resonance frequency, we observed increased attenuation for increased number of wirebonds, with frequency dependence and magnitudes in qualitative agreement with theory.  Loss from the circuit board was important above approximately 5\,GHz, as expected for an FR-4 substrate and copper metallization.  The predicted resonance frequency $f_{res}$ is shown as a vertical black line in all plots, which shows reasonable agreement between experiment and theory.

Panels A-D show that the resonance frequency $f_{res}$ increases with decreasing ground capacitance, as expected by theory.  Changing the wirebond spacing (panels E-F) and wirebond length (panels G-H) show a change in frequency in the correct direction.  The change is smaller than predicted for panels F and H, probably due to mutual inductance between the bonds.

\begin{figure}
\begin{center}
\includegraphics[width=3in]{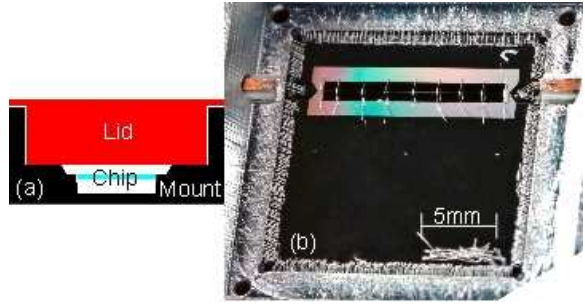}
\end{center}
\caption{Design of the chip mount. (\textbf{a}) Cross-sectional view of the mount, showing main aluminum box (black), chip (cyan), and lid (red). To raise $f_{res}$, we use cavities above and below the chip with height 1.4\,mm and 1.9\,mm, respectively. (\textbf{b}) Photo of chip in mount, showing numerous small wirebonds to ground around the edge of the chip to suppress crosstalk.  The wirebond spacing and length are approximately 0.13\,mm and length 0.7\,mm, respectively.  This chip has a coplanar resonator between two central pads, with wirebonds bridging the resonator to suppress slotline modes.}
\label{fig:boxdesign}
\end{figure}

With some confidence in the basic model, it is possible to extract useful information about practical design issues, such as the effect of nonuniform wirebonding, which occurs as a result of unsuccessful wirebonds.  Using SPICE simulations, we modeled a gap in the wirebonds with an increased inductance $L$ and capacitance $C$ at one location in the ladder.  Surprisingly, the lower-frequency resonance at this one location produces transmission through the entire ladder, leading to a peak in the transmission as seen in Figure \ref{fig:spice_res}. This indicates that wirebonds should be made as uniformly as possible all the way around the chip. However, note that even with a gap of $10$ bond spacings, the resonance frequency is only reduced by a factor of two. For the case of long wirebonds, the resonance frequency is lowered at one spot in the ladder, but does not lead to a significantly lowered resonance in the transmission.  In contrast, increasing the capacitance locally does lower the resonance frequency. In summary, small irregularities in wirebonding do not appear to be critical to the mount performance.

\section{Mount for large chips}

We used this model to design a new mount and to calculate its expected performance. To determine the effects of wirebond spacing and length, COMSOL was used to calculate the inductance and capacitance for a mount with different wirebond configurations. We assumed the mount had ground planes 1.4\,mm above and 1.9\,mm below the chip. We considered the effect of various bond spacings with a gap between the box and chip ground plane of $200\,\mu$m and a bond length of $400\,\mu$m, which accounts for the arching of the bond.  For a bond spacing of $500\,\mu$m, giving a total of 20 bonds per cm, we calculated $f_{res} = 43$\,GHz and $L/L_W = 0.37$, for an attenuation of 5.2\,dB per bond (104\,dB/cm). A bond spacing of $200\,\mu$m, for a total of 50 bonds per cm, gives $f_{res} = 68$\,GHz and $L/L_W = 0.15$, and an attenuation of 3.3\,dB per bond (167\,dB/cm). These results show that acceptable attenuation is possible using parameters that are experimentally feasible.

A new mount was designed for chips with 1\,cm sides to achieve these wirebonding parameters and to experimentally test performance.  As illustrated in Figure \ref{fig:boxdesign}, we used many short wirebonds between the chip and the mount to reduce crosstalk. Stray capacitance was minimized by designing a box with cavities both above and below the chip. The box was made from aluminum to shield stray magnetic fields when the devices are cooled to temperatures below 100\,mK.

\begin{figure}
\begin{center}
\subfigure{\includegraphics[width=3in]{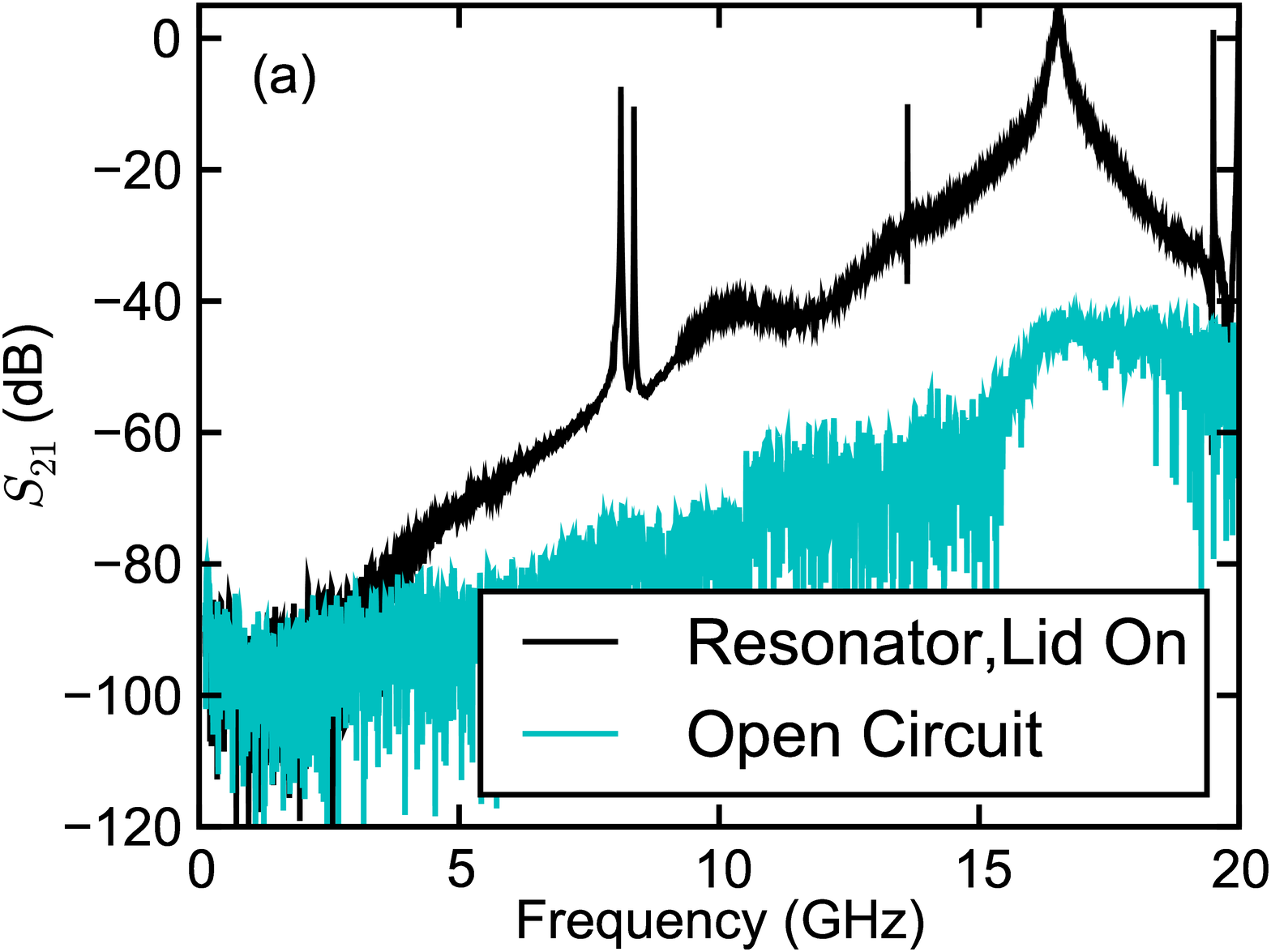}}
\subfigure{\includegraphics[width=3in]{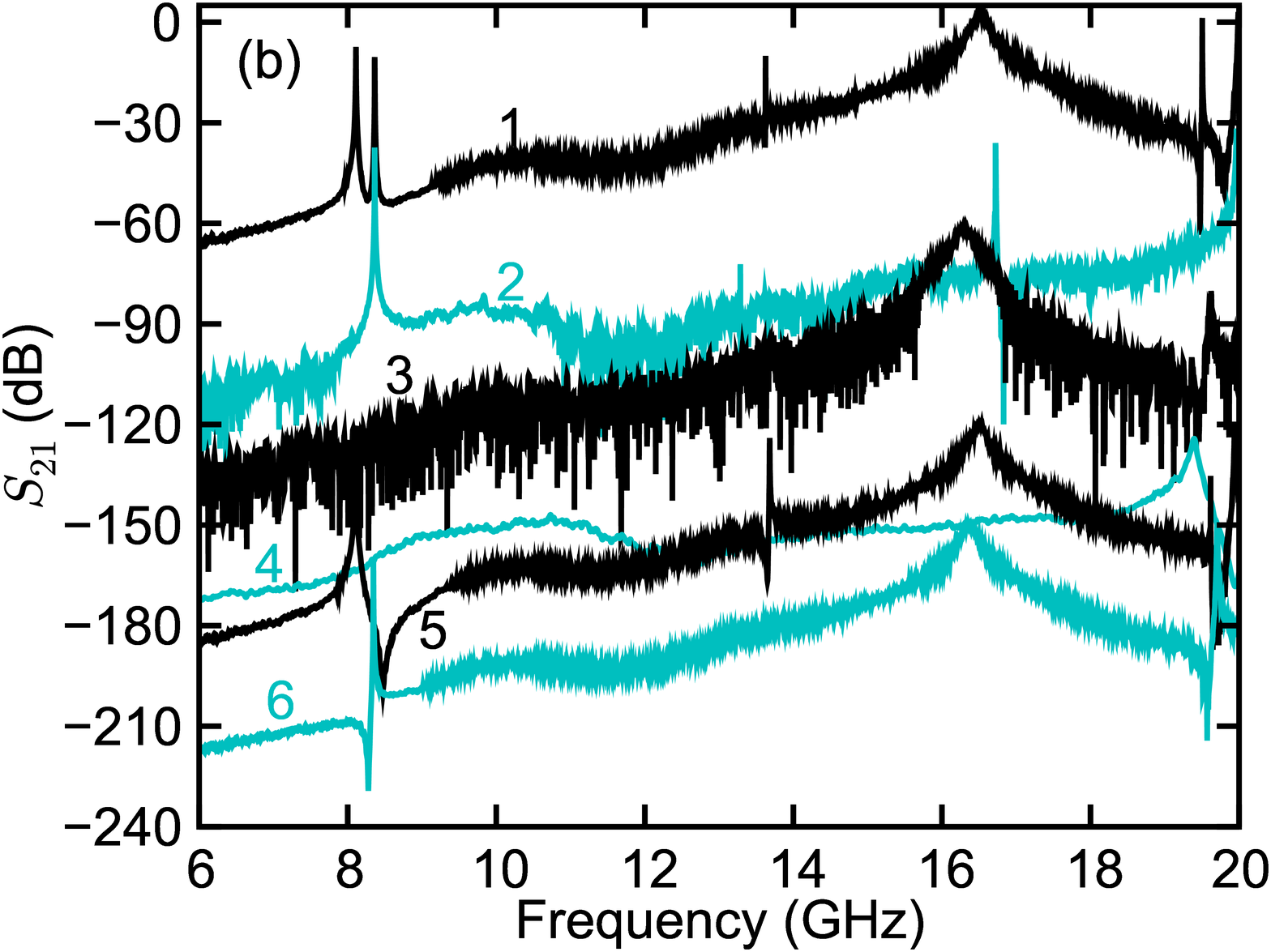}}
\end{center}
\caption{Identification of resonance modes. (\textbf{a}) Plot of mount transmission $S_{21}$ versus frequency for a coplanar resonator (black) and an open circuit without the mount (cyan), measured at 100\,mK.  The crosstalk is found to be -65\,dB at 6\,GHz, and somewhat less without the chip.  (\textbf{b}) Expanded scale of transmission showing major resonances, which can be identified by various test experiments.  Additional cases are each displaced by 30\,dB from each other for clarity. Case 1 is for the chip data in (a), and 2 is for the resonator with the lid off. Cases 3 and 4 are taken at room temperature, with the lid on and off, respectively. These four cases identify the cavity modes. Case 5 is for an identical chip design but with the center conductor removed, and 6 is for the original chip (case 1) after adding wirebonds that bridge the slotline gap.  The suppression of resonances with these shorts identifies the slotline modes.}
\label{fig:boxmodes}
\end{figure}

We used a coplanar resonator chip to quantify the mount performance.   The resonators were optically patterned 150\,nm thick aluminum films sputtered on silicon. The chip mount was cooled by an adiabatic demagnetization refrigerator to approximately 100\,mK.  The results from resonator measurements are plotted in Figure \ref{fig:boxmodes}, which have the transmission magnitudes normalized by removing cable attenuation in order to express the results as a mount response.

As illustrated in Figure \ref{fig:boxmodes}(a), we measured a transmission of -65\,dB at 6\,GHz, only somewhat larger than coming from open-circuit transmission without the mount.  The noise in the data comes from the noise floor of the network analyzer.

\section{Identification of spurious modes}

As shown in Figure \ref{fig:boxmodes}(b.1), we observed several resonance modes when the chip was measured at 100\,mK.  This is partly due to the large size of the mount, which has dimensions of 1\,cm, approximately a half-wavelength for the highest frequencies studied here.  In order to identify the modes, we performed several additional experiments.

One test was to measure the resonator at room temperature, as shown in Figure \ref{fig:boxmodes}(b.3) and (b.4). The presence of resonance modes at 16\,GHz and at 19.6\,GHz indicates these are cavity modes, as the lack of superconductivity precludes a chip resonance mode. As a further test, we then either removed the lid (see Figure \ref{fig:boxmodes}(b.2)) or placed Eccosorb LS-26 microwave-absorbing foam inside the sample mount lid (data not shown), and observed the disappearance of the 16\,GHz resonance leaving only the 19.6\,GHz peak.  We interpret this as the 16\,GHz peak coming from a cavity resonance between the chip and mount lid. The 19.6\,GHz peak probably comes from the smaller cavity below the chip.

Slotline modes are another source of undesirable resonances, which correspond to having unequal voltages on the ground planes on either side of the coplanar resonator. To identify these modes, a resonator chip was fabricated the same as the coplanar resonator but with no center trace; these measurements are plotted in Figure \ref{fig:boxmodes}(b.5), where we find resonance modes at 8.1\,GHz, 13.6\,GHz, and 20\,GHz.  Since these have not previously been identified as cavity modes, we identify them as slotline mode.  Note that their frequency ratios are close, but not equal, to 1:2:3.  These modes may be suppressed by airbridge wiring spanning the coplanar leads, which forces the voltages on either side to be equal. As illustrated in Figure \ref{fig:boxdesign}, we emulated this by spanning all coplanar traces with about 30 wirebonds, although we later found that 9 wirebonds sufficed. When doing so, we found these three slotline modes were all suppressed, as shown in Figure \ref{fig:boxmodes}(b.6).

We identify the mode at 8.4\,GHz as the coplanar resonator mode, since it is not a cavity or slotline mode.  This matches the design value of about 9\,GHz.  This was further confirmed by cooling through the transition temperature while applying a 2.5\,G magnetic field; as this value exceeds the critical field for the 6\,$\mu$m width traces for flux trapping \cite{Stan2004}, one expects a significant decrease in the $Q$ factor \cite{Song2009,Wang2009}. By following the approach in \cite{OConnell2008} for measuring $Q$, we found this mode dropped in $Q$ by a factor of 10, whereas other modes were only degraded by at most a factor of two.  We also found the characteristic field size for degradation of $Q$ was consistent with the width of the resonator's center trace.

Finally, when the lid was removed, an additional resonance mode was present at 16.7\,GHz, as shown in Figure \ref{fig:boxmodes}(b.2). With the lid attached, the 16\,GHz cavity mode had masked this additional mode. The $Q$ for this mode decreased by a more than a factor of 10 when field cooling, implying this is a coplanar mode.  This is consistent with expectations since the frequency of this mode is almost twice that of the fundamental coplanar mode.

We note that increasing the chip size beyond 1.5\,cm will lower the frequency of cavity modes to that near the qubit, and issue that will clearly have to be addressed in future research.  Since we observed that damping reduces cavity crosstalk in Figure \ref{fig:boxmodes}(b.4), we expect that efficient damping of these modes will enable the construction of large chip mounts with low crosstalk.

\section{Conclusion}

In conclusion, we have modeled crosstalk in a microwave chip mount as arising from non-zero wirebond inductance.  A impedance ladder model has been developed that can predict the performance on the grounding, and was experimentally verified with a scale model.  Our calculations indicate that stray transmission falls off at low frequencies with increasing wirebond density, and reaches near unity at a resonance frequency determined by the wirebond length and the stray capacitance between the chip and mount grounds. Hence, to improve the mount, it is necessary to use a high density of short grounding wirebonds and to decrease stray capacitance by using a mount without a ground plane under the chip.

A new chip mount was developed with this improved design in mind. We were able to characterize various resonances, including coplanar, cavity, and slotline modes.  The stray coupling was reduced to about -65\,dB at 6\,GHz.  This new mount is compatible with centimeter-sized chips for future qubit devices and points the way for even large chip sizes.

\end{document}